\documentclass{article}
\usepackage{spconf, amsmath, graphicx, siunitx, booktabs, fixltx2e, tabularx, xcolor}
\usepackage{hyperref}

\newcommand{\Expect}{{\rm I\kern-.3em E}}
\title{Boosting the rule-out accuracy of deep disease detection using class weight modifiers
\thanks{This paper was accepted by IEEE ISBI 2019. \copyright 2019 IEEE. Personal use of this material is permitted. Permission from IEEE must be obtained for all other uses, in any current or future media, including reprinting/republishing this material for advertising or promotional purposes, creating new collective works, for resale or redistribution to servers or lists, or reuse of any copyrighted component of this work in other works.}
}
\name{Alexandros Karargyris$^*$, Ken C. L. Wong$^*$, Joy T. Wu, Mehdi Moradi$^{**}$, Tanveer Syeda-Mahmood}
\address{IBM Research, Almaden Research Center - San Jose, CA  \\ \textit{$^*$Equal contribution, co-lead authors.} \\ \textit{$^{**}$corresponding author: mmoradi@us.ibm.com}}

\graphicspath{{./figures/}}

\begin{document}
%
\maketitle
\begin{abstract}

In many screening applications, the primary goal of a radiologist or assisting artificial intelligence is to rule out certain findings. The classifiers built for such applications are often trained on large datasets that derive labels from clinical notes written for patients. While the quality of the positive findings described in these notes is often reliable, lack of the mention of a finding does not always rule out the presence of it. This happens because radiologists comment on the patient in the context of the exam, for example focusing on trauma as opposed to chronic disease at emergency rooms. However, this disease finding ambiguity can affect the performance of algorithms. Hence it is critical to model the ambiguity during training. We propose a scheme to apply reasonable class weight modifiers to our loss function for the no mention cases during training. We experiment with two different deep neural network architectures and show that the proposed method results in a large improvement in the performance of the classifiers, specially on negated findings. The baseline performance of a custom-made dilated block network proposed in this paper shows an improvement in comparison with baseline DenseNet-201, while both architectures benefit from the new proposed loss function weighting scheme. Over 200,000 chest X-ray images and three highly common diseases, along with their negated counterparts, are included in this study.

\end{abstract}
\section{Introduction}
\label{sec:intro}

Chest X-rays (CXR) are one of the most commonly performed medical imaging exams as part of the initial diagnostic workup and screening processes in various clinical settings, ranging from the primary care office to the emergency room department. Given the widespread use of this modality, automation of the CXR read has become an important goal of the medical imaging community \cite{Conference:Wang:CVPR2017,TexRay}. These efforts have been aided by the publication of two fairly large public datasets \cite{Journal:Johnson:arXiv2019,NIHcxr_wang2017chestx}.

The use of these image datasets in classifier training required natural language processing to extract the training labels from the CXR radiology reports to build labeled datasets at scale. Viewing any CXR radiology report, one would soon notice three main categories for each finding: 1) finding label was affirmed/positive, 2) finding label was negated, and 3) finding label was not mentioned in the report. In fact, because CXR is often used as a screening exam to rule out abnormal findings, a large number of sentences in most reports would specifically mention that some findings are not present (negated). An example would be “no pneumothorax, pleural effusion and consolidation”. Therefore, directly predicting a negated finding output for some findings could be just as useful clinically as a positive finding prediction because the information is still helpful in guiding the subsequent patient management.

However, it is less certain how one should deal with the “no mention” category for all the different finding types, which represent an even larger proportion of the whole label space. Given the wide clinical applications of CXRs and over a hundred different types of findings of varying prevalence, there are multiple reasons why any finding might not be mentioned in the radiology report. The “no mention” cases could be due to: 1) true negation: the finding label is not present but also clinically not important enough to specifically negate in report, or 2) false negative: the finding is present but the radiologist missed it or did not think it was clinically relevant enough to mention in that particular setting (e.g. reporting an irrelevant chronic finding like shoulder arthritis in an acute trauma case).

Therefore, adjusting the training of classifiers built using NLP-labelled image collections to deal with the “no mention” cases is a key problem to address to avoid discarding a large amount of CXR examples that were essentially “partially” labeled. To inform our experiments, we did a manual exploratory analysis of three open source CXR datasets \cite{NLMIndiana_demner2015preparing, PLCO_gohagan2000prostate, NIHcxr_wang2017chestx} and a radiology literature review of missed findings in CXRs. The exploratory analysis of 150 random CXR images, where the NLP labels were “no findings” or “normal”, showed that 6\% of the images had an abnormal anatomical finding. The literature search suggested that lung nodules and fractures have a missed finding rate between 20-40\% amongst patients who eventually had the finding confirmed \cite{quekel1999miss,de2015pitfalls,harris1988prevalence}.

This paper explores a solution to optimize the training of a disease finding classifier in situations where both positive and negated labels are present, by addressing the uncertainties of the no mention cases in each label. We propose a scheme to apply reasonable class weight modifiers to our loss function for the no mention cases during training. Our clinically guided hypothesis is that if a positive finding is not mentioned in the report, we are more certain that it is a true negative. On the other hand, if a negated finding is not mentioned, it has a higher chance to be a false negative. Therefore, in our scenario, we propose a higher weight to the no mention case for the positive label than for the corresponding negated label to reduce the influence of the ambiguous no mention cases.

We train two different deep neural net architectures and show that in both cases the use of these class weight modifiers with the loss function result in improved accuracy in detection of negative labels. The studied architectures are DenseNet and a custom network architecture reported in this paper for the first time.

\section{Methods}
\label{sec:methods}

In order to build a deep neural network for producing findings necessary to compose a CXR report, we needed a very large number of labeled images. We achieved this by automatic text analysis of the reports accompanied by the MIMIC-CXR dataset \cite{Journal:Johnson:arXiv2019}. In this paper, we mostly discuss the process of building the finding classifier and the novel loss function and architecture proposed here, with only a brief description of the text analysis methodology that produced the labels from text reports.

\subsection{Label extraction}
A top-down knowledge-driven plus a bottom up text curation process was used to identify a set of unique finding concepts relevant for CXRs. We also used an NLP concept expansion engine \cite{coden2012spot} to semantically map the different ways a finding could be described in reports to a discrete finding label set validated by radiologists. We then applied context recognition NLP to differentiate between negated and affirmed instances for each finding mention \cite{syeda2015learning}. Where CXR reports did not mention a finding, it is flagged as a "no mention" case. We picked the three most frequently occurring finding labels, and their negated versions, to conduct the experiments in this study.

\subsection{Class weights and loss function}
For each semantic label, the numbers of positive and negated samples can be highly unbalanced, and the class with the higher frequency can dominate the loss function and lead to suboptimal classifiers. Therefore, class weights are usually used to alleviate this issue. We compute the class weights as:
\begin{equation}\label{eq:class_weights}
  w_1 = \frac{f_0}{f_1 + f_0},\ \ \ w_0 = 1 - w_1
\end{equation}
with $w_1$ and $w_0$ the weights for the positive and negated classes, and $f_1$ and $f_0$ the numbers of the positive and negated samples, respectively. The loss of each semantic label can then be computed as the weighted binary cross-entropy:
\begin{equation}\label{eq:loss_func}
  L = w_1(- y \ln(p)) + w_0 (1 - y)(-\ln(1 - p))
\end{equation}
where $y$ = 1 for positive samples and 0 otherwise. $p \in [0, 1]$ is the sigmoid output from the network prediction. The average loss of all semantic labels is used for the backpropagation.

\begin{table}[t]
\small
\centering
\begin{tabular}{lll}
\toprule
A & $\bar{\mathrm{A}}$ & Meaning \\
\midrule
1 & 1 & Contradiction \\
1 & 0 & A exists \\
0 & 1 & $\bar{\mathrm{A}}$ exists \\
0 & 0 & Ambiguous \\
\bottomrule
\end{tabular}
\caption{Meanings of different combinations of a negated pair (a semantic label A and its negation $\bar{\mathrm{A}}$).}
\label{table:logic}
\end{table}

\subsection{Class weight modifiers}

With the introduction of negations in the semantic labels, the interpretation of a sample with both negatives (0, 0) for a pair (a semantic label and its negation, e.g. "consolidation" and "no consolidation") can be ambiguous. Table \ref{table:logic} shows the possible combinations of a negated pair. For a semantic label, as the positives (1's) are explicitly mentioned by radiologists, they are certain findings. On the other hand, the negatives (0's) are not mentioned and can be ambiguous, because apart from the negative meaning of the semantic label, the 0's can also mean the finding is missed or not considered. For example, for the negated label "no consolidation", a 0 can mean there is consolidation, or "no consolidation" is not considered at all. Therefore, the (1, 1) pair is contradicting and should not exist, the (1, 0) and (0, 1) pairs should follow the meanings of 1's as they are conscious annotations, and the (0, 0) pair is ambiguous.

To handle such ambiguity in training, we propose the weight modifiers to modify the class weights of each sample with the (0, 0) negated pair when computing the loss function. In fact, although 0's are ambiguous in general, the level of ambiguity is different between a semantic label and its negation. For findings such as "consolidation", the chance of being missed or not considered should be low as radiologists are trained to report anomalies. For the negations such as "no consolidation", the chance of being not considered is high as radiologists are usually not required to explicitly mention non-existence of all findings. Therefore, the weight modifiers for a semantic label ($m$) and its negation ($\bar{m}$) are different, which are given as:
\begin{equation}\label{eq:modifiers}
  m = N(\mu, \sigma),\ \ \ \bar{m} = 1 - m
\end{equation}
with $N(\mu, \sigma)$ the normal or Gaussian distribution with mean $\mu$ and standard deviation $\sigma$. $m$ and $\bar{m}$ are multiplied by $w_0$ in (\ref{eq:class_weights}) during training. A larger $m$ means we trust a semantic label more than its negation, and vice versa. Instead of a constant $m$, a normal distribution is used to model the uncertainties caused by ambiguity. $\sigma$ was fixed at 0.05 in the experiments, while different values of $\mu$ were investigated.

\begin{figure}[t]
\begin{minipage}[b]{1.0\linewidth}
  \centering
  \centerline{\includegraphics[width=9cm]{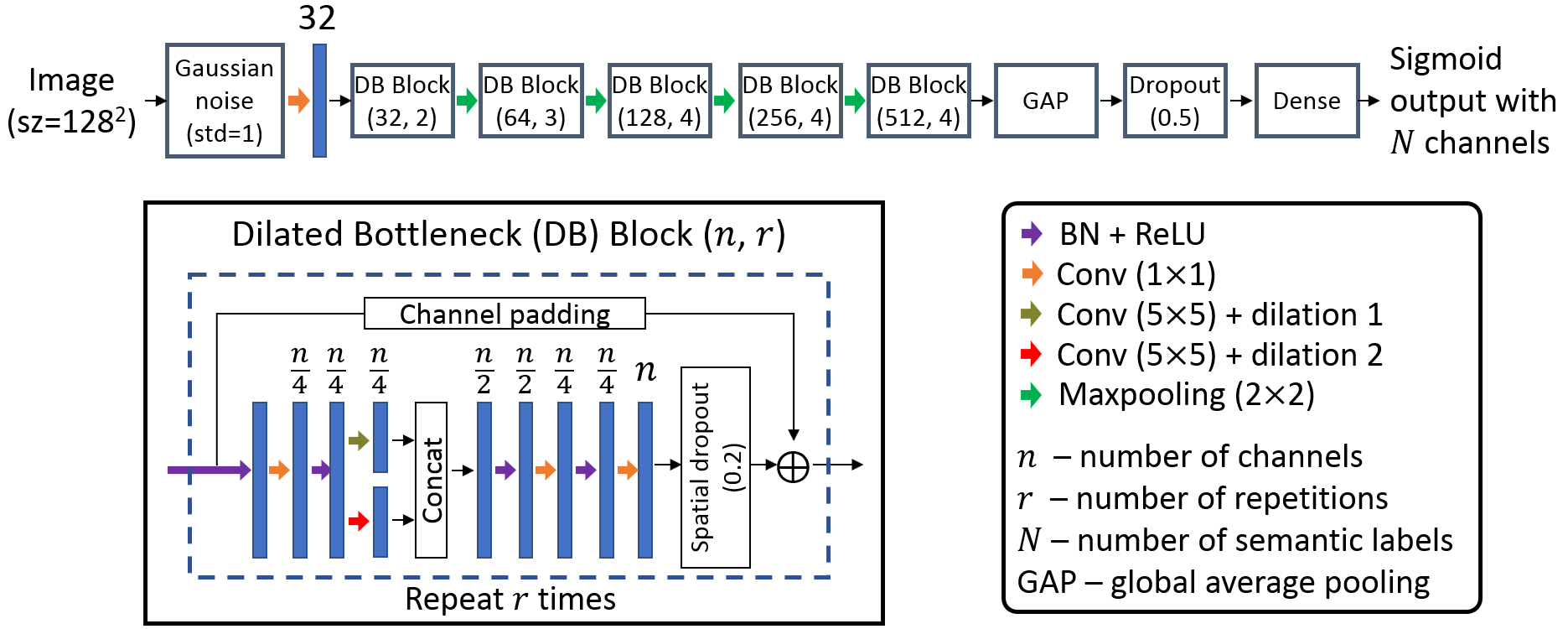}}
\end{minipage}
\caption{Custom network architecture used in this work.}
\label{fig:custom_net}
\end{figure}

\subsection{Network architectures}

To show that the proposed weight modifiers are generally applicable, we performed experiments on a custom architecture, and also on a widely used architecture DenseNet \cite{huang2017densely}.

The custom architecture comprises the proposed Dilated Bottleneck (DB) blocks (Fig. \ref{fig:custom_net}). In each block, the efficient bottleneck architecture of ResNet is used so that deeper network can be trained \cite{Conference:He:CVPR2016}. Dilated convolutions with dilation rates of 1 and 2 are also used to aggregate multi-scale context \cite{Journal:Yu:arXiv2015}. Identity mappings and pre-activations are used for better information propagation \cite{Conference:He:ECCV2016}, and spatial dropouts with dropout probability of 0.2 are used to alleviate overfitting \cite{Conference:Tompson:CVPR2015}. Therefore, each block allows efficient learning of multi-scale information. To further alleviate overfitting, a Gaussian noise layer, global average pooling, and dropout with probability of 0.5 are used with the cascaded DB blocks to form the network architecture. Images are resized to 128$\times$128 with this architecture.

We also used DenseNet \cite{huang2017densely} for the same problem to show the improvements from modifiers can be repeated on other networks. DenseNet utilizes skip connections to feed information to latter layers. We used DenseNet with 201 layers, and 18,319,554 trainable parameters.

\begin{table*}
  \centering
\caption{The break-down of images across labels }
\label{table:break_down}

\begin{tabular}{|c|c|c|c|c|c|c|}
\hline
 & Consolidation & No  & Pneumothorax & No  & Pulmonary  & No Pulmonary \\

  &  & Consolidation  &  & Pneumothorax & Edema &  Edema\\

\hline
Training & 12,088 & 57,920  &2,443 & 107,750&15,517& 32,967\\

\hline
Validation & 1,677 &8,278  &397 & 15,474 & 2,226 & 4,576 \\

\hline
Testing & 3,413 & 16,461  & 709 & 30,711 & 4,449 & 9,317\\
\hline
\end{tabular}
\end{table*}

\subsection{Training strategy}

Image augmentation with rigid transformations is used to avoid overfitting. As most of an image should be included, we limit the augmentation to rotation ($\pm$10 degrees), shifting ($\pm$10\%), and scaling ([0.95, 1.05]). The probability of an image to be transformed is 80\%. The optimizer Adam is used with a learning rate of 10$^{-4}$, a batch size of 64, and 20 epochs.

We are using an IBM POWER9 Accelerated Computer Server (AC922) that is designed to accommodate the data-intensive characteristics of modern analytics and AI workloads by fully exploiting its GPU capabilities, eliminating I/O bottlenecks and sharing memory across GPUs and CPUs. The machine is equipped with four V100 NVidia GPUs in its air-cooled configuration which we used.

As a proof of concept, six semantic labels of three negated pairs ("consolidation", "no consolidation"), ("pneumothorax", "no pneumothorax"), and ("pulmonary edema", "no pulmonary edema") were used, resulting in 204k frontal chest X-ray images. The choice of these pairs was intentional because these pairs have a high frequency in the MIMIC-CXR dataset and thus made our experiments statistically safe. The break-down of samples is listed in Table \ref{table:break_down}.

The dataset was divided into 70\% for training, 10\% for validation, and 20\% for testing, and the testing results are reported. Different values of $\mu$ in (\ref{eq:modifiers}) were investigated. A value of 0.9 means we trust a semantic label more than its negation, and a value of 0.1 means the opposite. Note that while all possible sample combinations are included in the training phase, at the time of testing, we only test on samples that are not ambiguously labeled so we can measure the performance changes without ambiguity.

\section{RESULTS}

Our first observation is that a large number of cases in MIMIC-CXR radiology reports contained ambiguous disease findings (e.g. 50\% ambiguous consolidation cases, 23\% ambiguous pneumothorax cases, 66\% ambiguous pulmonary edema cases). This shows the importance of modeling the ambiguity of labels during training.

\begin{table*}
  \centering
\caption{Baseline and improved performance, in terms of area under ROC curve, of the two architectures. Note the marked improvement of the No Pulmonary and No Pneumothorax labels when the ambiguity modeling is performed with appropriate weight values. }

\label{table:performance}
\small
\begin{tabular}{|c|c|c|c|c|c|c|}
\hline
	  & Consolidation & No  & Pneumothorax & No  & Pulmonary  & No Pulmonary \\

  &  & Consolidation  &  & Pneumothorax & Edema &  Edema\\  \hline
	Dilated Block, baseline & 0.83 & 0.82 & 0.81 & 0.69 & 0.87 & 0.80 \\
	
	Dilated Block, $\mu=0.8$ & 0.82 & 0.83 & 0.81 & \textbf{0.80} & 0.87 & \textbf{0.87} \\
	
	 \hline
	 \hline
	  DenseNet-201, baseline & 0.83 & 0.83 & 0.81 & 0.72 & 0.87 & 0.81 \\
	
	DenseNet 201 $\mu=0.9$ & 0.83 & 0.84 & 0.82 & \textbf{0.80} & 0.88 & \textbf{0.87 }\\
	
\hline
\end{tabular}
\end{table*}

\begin{figure}[t]
\begin{minipage}[b]{1.0\linewidth}
  \centering
  \centerline{\includegraphics[width= 7 cm]{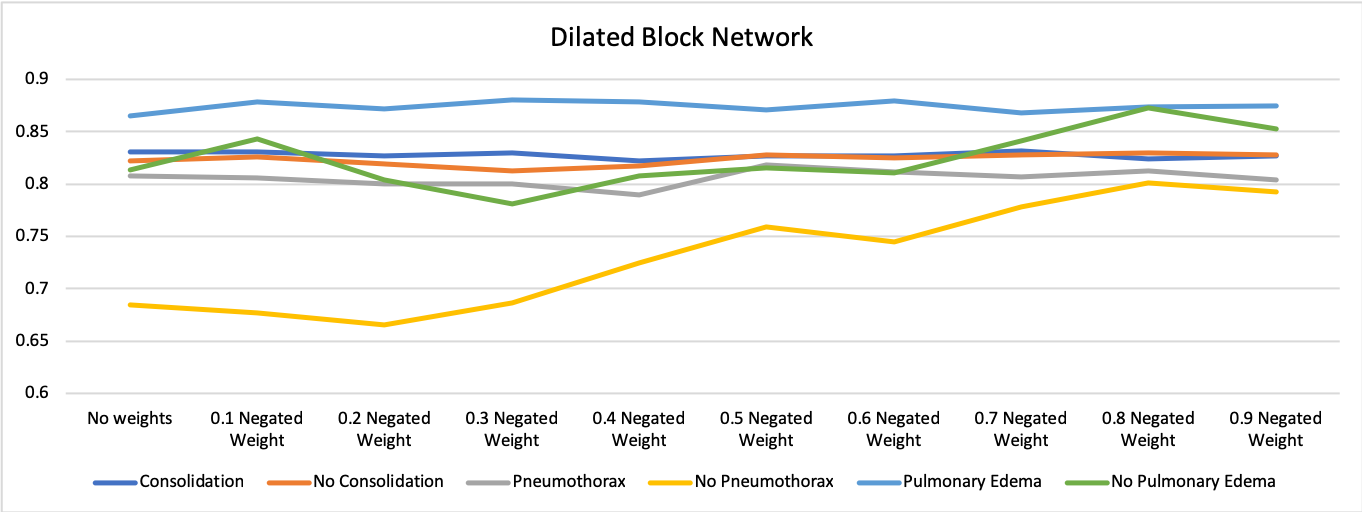}}
\end{minipage}
\caption{Area under ROC curve for dilated block network architecture for varying values of weight modifier for negated findings.}
\label{fig:dilated_roc}
\end{figure}

\textbf{Dilated block network: } The baseline performance of Dilated block net on the six labels, along with the performance at best weight combination in the proposed method are reported in the Table \ref{table:performance}. Fig. \ref{fig:dilated_roc} depicts the ROC per label for all combinations of weights for $\mu=0.1$ to $\mu=0.9$. The optimal weight was $\mu=0.8$, chosen based on average area under ROC curve for all six findings. The improvement is primarily on the negated labels as expected. The area under ROC curve for no pneumothorax increases from 0.69 to 0.80, and from 0.80 to 0.87 for no pulmonary edema. The performance change for no consolidation is smaller.

\begin{figure}[t]
\begin{minipage}[b]{1.0\linewidth}
  \centering
  \centerline{\includegraphics[width=7 cm]{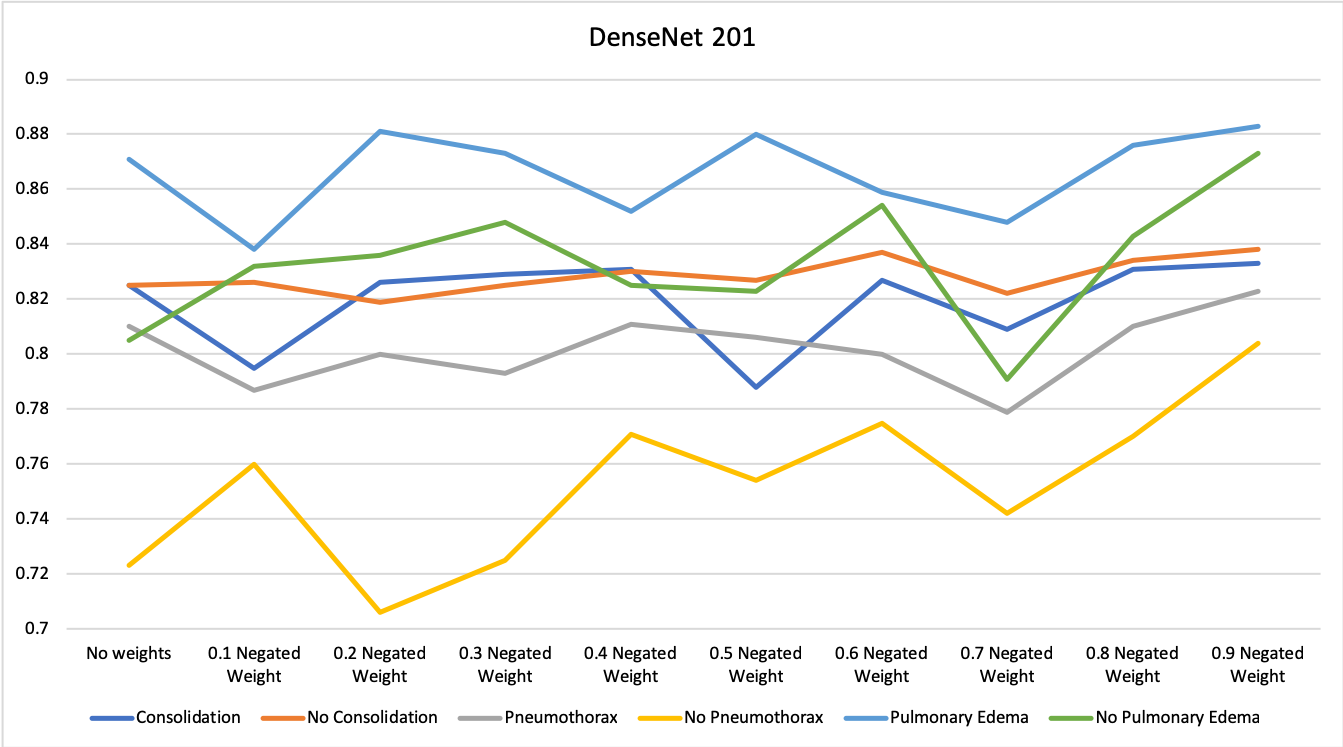}}
\end{minipage}
\caption{Area under ROC curve for DenseNet-201 architecture for varying values of weight modifier for negated findings.}
\label{fig:dense_roc}
\end{figure}

\textbf{DenseNet Results: }DenseNet results are in Fig. \ref{fig:dense_roc} and the second half of Table \ref{table:performance}. The optimal weight was $\mu=0.9$. Again the improvement is primarily on the negated with the area under ROC curve for no pneumothorax increasing from 0.72 to 0.80, and from 0.81 to 0.87 for no pulmonary edema. The performance stays similar to baseline for positive findings.

\begin{figure}[t]
\begin{minipage}[b]{1\linewidth}
  \centering
  \centerline{\includegraphics[width=8 cm]{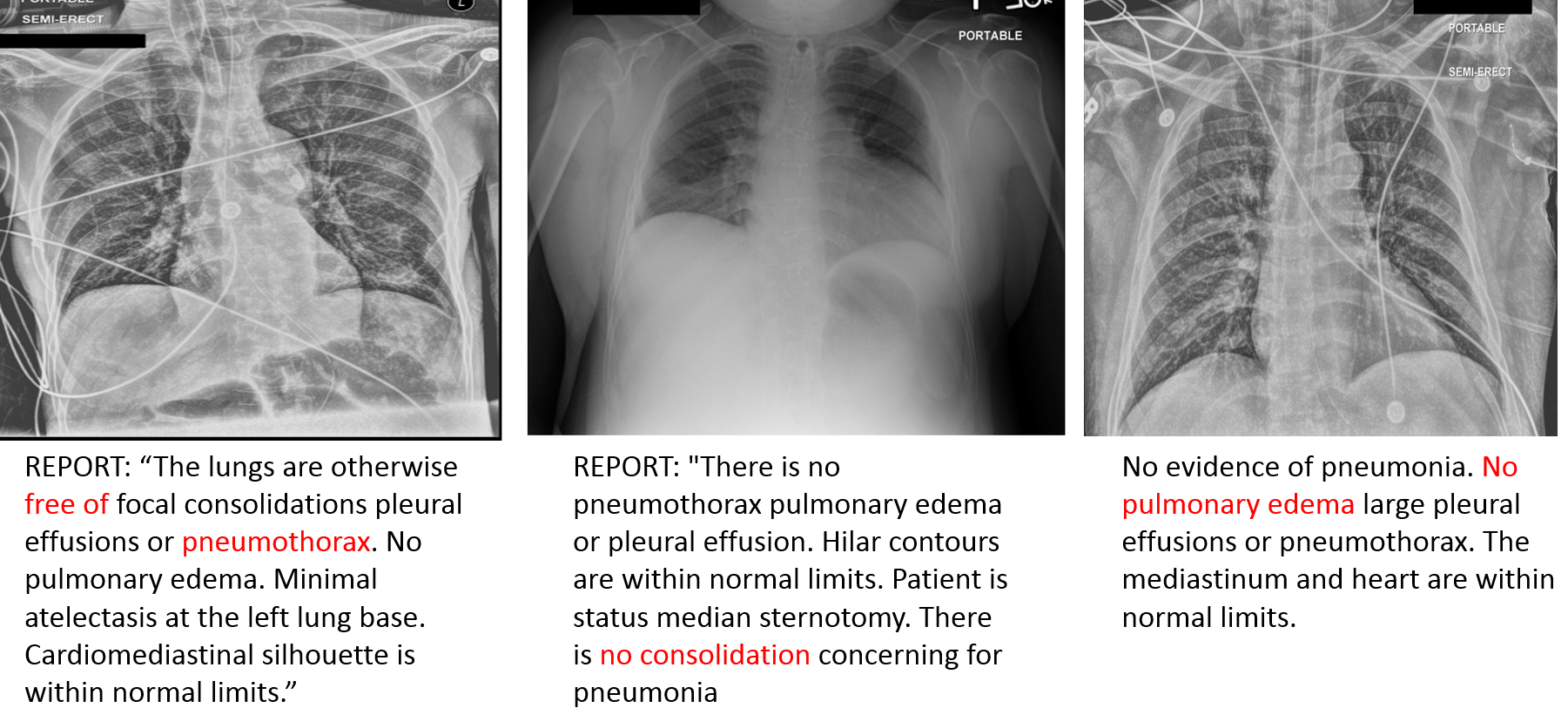}}
\end{minipage}
\caption{Examples of misclassification on baseline DenseNet. In all these cases, the DenseNet model trained with the proposed modifiers correctly classified the image as negated for the highlighted finding}
\label{fig:allcases}
\end{figure}

\textbf{
Examples of corrections: }Since the test set consists of only non-ambiguous labels, the performance improvement translates to objectively more accurate findings. Nevertheless, for illustration purposes, Fig. \ref{fig:allcases} shows three examples within the test set of cases where the use of the proposed weight modifiers at the time of training has changed the prediction from false positive to true negative. The text under each image is the original report written by the radiologist and the negated finding of interest is highlighted.

\subsection{Conclusions}

In this paper we presented a methodology to deal with ambiguity of disease findings in radiology reports. Our approach to model this ambiguity is to add a class weight modifier and evaluate a range of weights from 0.1 to 0.9 for impact on classification accuracy in non-ambiguous test cases. The optimal balance of probabilities is that 80-90\% of the ambiguous cases are negated disease findings. This was verified by two independent state-of-the-art neural networks evaluated on many images. We observed a large improvement in negated disease findings classification on a very large dataset while maintaining similar levels of accuracy in positive disease findings. This work builds toward our efforts of building a radiologist assistant application that can perform similar to an entry level radiologist and prepare the first version of a radiology read for CXR.

\newpage
\bibliographystyle{IEEEbib_init}
\bibliography{Ref}

\end{document}